\documentstyle[psfig]{article}
\begin{document}
\renewcommand{\theequation}{\thesection.\arabic{equation}}
\def\bq{\begin{equation}}
\def\eq{\end{equation}}
\def\bqa{\begin{eqnarray}}
\def\eqa{\end{eqnarray}}
\def\RR{\hbox{\it I\hskip -2.pt R }}

\vspace{2cm}

\title{Critical properties of $\phi^4_{1+1}$-theory in
Light-Cone Quantization}
\author {St\'ephane SALMONS$^{(1)}$, Pierre GRANG\'E$^{(1)}$
and Ernst WERNER$^{(2)}$\\
(1)Laboratoire de Physique Math\'ematique et Th\'eorique\\
UMR 5825, Universit\'e Montpellier II \\
F-34095, Montpellier, Cedex 05 - FRANCE \\
(2)Institut f\"ur Theoretische Physik\\
D-93040, Universit\"at Regensburg - GERMANY}
\date{}
\maketitle

\vspace{2.5cm}

\begin{abstract}
The dynamics of the phase transition of the continuum $\Phi ^{4}_{1+1}$-theory
in Light Cone Quantization is reexamined taking into account fluctuations of
the order parameter  $< \Phi >$ in the form of dynamical zero mode operators
(DZMO) which appear in a natural way via the Haag expansion of the field $\Phi (x)$ of the interacting theory. The inclusion of the DZM-sector changes significantly 
the value of the critical coupling, bringing it in agreement within $2 \%$ with the
 most recent Monte-Carlo and high temperature/strong coupling estimates. The critical slowing down of the DZMO governs the low momentum behavior of the dispersion
relation through invariance of this DZMO under conformal transformations  preserving the local light cone structure. The critical exponent $\eta$ characterising the scaling behaviour at 
 $k^2 \rightarrow 0$ comes out in agreement with the known value $0.25$ of the Ising universality class. $\eta$ is made of two contributions: one, analytic $(75 \%$) and another ($25 \%$) which can be evaluated only numerically with an estimated error of $3 \%$. The $\beta$-function is then found from the non-perturbative expression of the physical mass. It is non-analytic in the coupling constant with a critical exponent $\omega=2$. However, at $D=2$, $\omega$ is not 
parametrisation independent with respect to the space of coupling constants due to this strong non-analytic behaviour.

\end{abstract}

\vfill
\hrule\vskip10pt
\noindent{
}

\setcounter{equation}{0}

\section{Introduction}

Light cone quantization (LCQ) of scalar field theory in two dimensions $(LC -
\phi^4_{1+1})$ has been studied rather extensively during the last years.
The early attempts  \cite{PaBr} used discretization in the space direction thereby dealing 
with the problematic zero mode problem of the field operators in a simple way. Besides its
interests as a testing ground for techniques in LCQ such scalar field theory is mainly
useful for its phase transition properties, well known in the context of equal time
quantization (EQT), and relevant to the description of many physical systems.

In LCQ it is recognized now that the phase transition is driven by a coupling between two
distinct parts of the field operator \cite{Hou}: the static zero mode and quantum fluctuations of
the particle field. The zero mode carries essential non-perturbative informations which in
EQT are present in the non-trivial vacuum structure. It is important to realize that
critical behaviour studies, were it in LCQ or EQT, can only be achieved for effective
theories in the  continuum. Hence in the
discretized versions the approach to the continuum limit must be under control, which may be  
problematic \cite{SaGrWe} .
Recently an LCQ- formulation in the continuum (CLCQ)\cite{GrUlWe} was proposed in which both the zero
mode and renormalization problems are treated in a consistant way. It is based on the
treatment of fields as operator-valued distributions. The critical coupling $g_c$ and critical 
exponents were calculated to lowest order in the Haag expansion of the field operators. 
Whereas for first order perturbation theory  
$g_c$ comes out $45\%$ below the  value obtained from Monte Carlo or high temperature estimates , it is only $20\%$ below that value for lowest order CLCQ. However
critical exponents are usually those of mean field theory.

To improve this situation we extend here the theory to second order in the Haag
expansion of the fields. The benefit is a very natural inclusion of fluctuations of the
order parameter. Clearly they are essential to discuss the dispersion relation in the vicinity
of the phase transition, a feature closely related to the
dynamics of the phase transition and to the critical behaviour. The following results
are obtained\\

1) - The critical coupling is significantly improved and comes out sligthly below the
canonical value.\\

2)- Non-trivial values of the critical exponents are related to the existence of
dynamical zero modes transfering momentum and not considered so far. They enter both in
the equation of motion of the classical field $\phi_c(x)$ and of the particle field
$\varphi_2(x)$, thereby leading to a distinctive dispersion relation.\\

In Section 2 the formalism of Haag series in second order is introduced and the zero
mode concept is extended to a dynamical operator able to transfer momentum. From the
equation of motion (EM) for the total field $\Phi(x)$ matrix elements of different
types are taken giving coupled equations for the different components of $\Phi(x)$. The
solution of these equations is presented in Section 3, leading to a transcendental
equation for the critical coupling. Its solution is discussed and solved
hereafter numerically. In Section 4 the $k^2 \to 0$ limit of the dispersion relation is
discussed at some length, to extract in a transparent way its specific feature leading
to a non-trivial value of the critical exponent $\eta$ (ie $\lim_{k \to 0} \Gamma_2(k)
\sim k^{2 - \eta})$. Finally our results for $\beta$-functions, critical coupling and
critical exponent $\omega$ are collected  and in Section 5 a general discussion and conclusions
are also given.

\setcounter{equation}{0}

\section{Haag series, equations of motion}

\subsection{Haag series in second order and dynamical zero modes}

The Haag series of the total field $\Phi(X)$ is written in terms of normal ordered
products of the free field $\varphi_0(X)$ with LC coordinates $(X^-, X^+)$ \cite{Hou}

\begin{equation} \Phi(X) = \phi_c(X) + \Psi(X), \end{equation}

with 

\begin{eqnarray}  
\Psi(X^-,X^+) & = & \varphi_0(X^-,X^+) + \int dX^-_1 dX^-_2 
:\varphi_0(X^-_1,X^+) \varphi_0(X^-_2,X^+):  \nonumber \\
&    & g_2(X^-_1-X^-, X^-_2 - X^- ; X) + ... \nonumber \\
& : = & \varphi_0(X^-,X^+) + \varphi_2(X^-, X^+) + \mbox {higher order terms }\\
&    &\mbox { in the number of free fields operators $\varphi_0$}. \nonumber
\end{eqnarray}

In Eq. (2.1) $\phi_c(x)$ and $\Psi(X)$ are respectively the classical and quantum parts of
the total field $\Phi(X)$. Due to the interaction, $\phi_c(X)$ and $\Psi(X)$ are coupled to
each other. This coupling is made explicit by the X-dependence in the wave function
$g_2(X^-_1 - X^-, X^-_2-X^- ; X)$. It will be seen more clearly later on when the coupled
equations for $g_2$ and $\phi_c(X)$ are written down. It should be noticed that the contributions 
$\varphi_2(X)$ and $\varphi_3(X)$ are of different character: while $\varphi_3(X)$ has perturbative
contributions, $\varphi_2(X)$ is purely non-perturbative and is only due to fluctuations of the 
order parameter. 

In CLCQ the free field $\varphi_0(X)$ is an operator valued distribution \cite{GrUlWe} with a
generalized Fock expansion built out of $C^{\infty}$-test functions  with compact support in $\RR^2$ . 
It is written as 

\begin{eqnarray}
\varphi_0(X) & = & \varphi^-_0(X) + \varphi^+_0(X) \nonumber \\
& = & \int {dk^+ \over {\sqrt{4 \pi k^+}}} [a(k^+) e^{{i \over 2}k.X} + a^+(k^+) e^{-{i
\over 2} k.X}] \theta(k^+) \hat{f}(k^+), 
\end{eqnarray}

where $\theta(k)$ is the usual step function and the creation and annihilation operators $a^+(k^+)$ and $a(k^+)$ obey  $[a(k^+),
a^+(k'^+)] = \delta(k^+ - k'^+)$, $k.X = \hat{k}^-X^+ + k^+X^-$, $\hat{k}^- = {m^2 \over k^+}$, 
 $(m^2 > 0)$. The properties of the test function $\hat{f}(k^+)$ are discussed in \cite{GrUlWe} :
its effect is to render the integral finite at both ends of the spectrum in $k^+$ 
and the parameter entering any possible form of $\hat{f}(k^+)$
is related in an essential way to the final renormalization procedure of the theory.

Written as in Eq.(2.2) $\varphi_2(X)$ is not apparently covariant since the
integrations are done on a fixed time sheet. However manipulations on
$\theta$-functions, standard in Eq.(2.3), can bring it into a manifestly covariant form
exhibiting advanced and retarded parts of the amplitude $g_2$.

The form (2.3) is used in Eq.(2.2) and a Fourier transformation with respect to $X^-_1-X^-$
and $X^-_2 - X^-$ is then performed, giving

\begin{eqnarray}
\varphi_2(X) & = & \int{{dk^+_1 dk^+_2 \over \sqrt{k^+_1 k^+_2}}} \theta(k^+_1) \theta(k^+_2)
\hat{f}(k^+_1) \hat{f}(k^+_2) \nonumber \\
&  & \{ g^{++}_2 (k_1,k_2 ; X) [a^+(k^+_1) a^+(k^+_2) e^{-{i \over 2}(k_1+k_2).X} +
a(k^+_1)a(k^+_2) e^{{i \over 2}(k_1+k_2).X}] \nonumber \\
& + & [g^{+-}_2 (k_1, -k_2 ; X) + g^{-+}_2(-k_2, k_1 ; X)] a^+(k^+_1) a(k^+_2) e^{{i \over
2}(k_2-k_1).X} \}. 
\end{eqnarray}

Here $k_i = (k^+_i, \hat{k}^-_i)$ with $\hat{k}^-_i$ defined after Eq.(2.3). Because of
hermiticity the coefficient function $g^{--}_2$ of $ a(k^+_1) a(k^+_2)$ is identical to
$g^{++}_2$. Hereafter the combinaition $g^{+-}_2(k_1, -k_2 ; X) + g^{-+}_2(-k_2, k_1 ; X)$
always occurs in this form and will be abbreviated by $G_2(k_1,- k_2 ; X)$.
{The zero mode of the field operator $\varphi_2$, introduced in \cite{HKW} where it is called
$\hat{\Omega}$, is contained in the form (2.4). It is obtained after integration over $X^-$ with
an invariant measure which, if momenta are measured in units of the renormalized mass $\mu$ as 
in \cite{GrUlWe} (see below) is simply ${\mu \over 4\pi} dX^-$. In this case, with $\bar{G}_2$ denoting the
$X^-$-independant part of $G_2$, the specification is just

\newpage

\begin{eqnarray}
{1 \over \sqrt{k^+_1k^+_2}} G_2(k_1,-k_2 ; X) |_{ZM} & = & {\mu \over k^+_1} \bar{G}_2(k_1,
-k_1) \delta(k^+_1 - k^+_2) \nonumber \\
& : = & C_0(k^+_1) \delta(k^+_1 - k^+_2).
\end{eqnarray}

It is important to realize that the zero mode concept - i.e. an $X^-$-independent field
contribution - can be extended to a dynamical operator transfering momentum. The physics
behind this process is as follows : the coefficient function 
$G_2(k_1,-k_2 ; X)=g^{+-}_2 (k_1, -k_2 ; X) + g^{-+}_2(-k_2, k_1 ; X)$ in Eq.(2.4)
can take a periodic space dependence if the particles propagate in a periodic background
field, the field $\phi_c(X)$ in this case. Depending upon the sign of $k^+_2 - k^+_1$ this
periodic evolution is of the form  $\exp[\pm {i \over 2} k^+X^-]$ and compensates the
periodic factor $\exp[{i \over 2} (k^+_2 - k^+_1)X^-]$ coming from momentum transfer.
Integrating Eq.(2.4) over $X^-$ in this case generates a particular type of zero mode,
which we note $\hat{\Omega}_k$. It takes the general form

\begin{equation}
\hat{\Omega}_k = \int dk^+_1 C_1(k^+_1, k^+) a^+(k^+_1 + k^+) a(k^+_1), 
\end{equation}

where the coefficient $C_1(k^+_1, k^+)$ is related to the coefficient function 
$\tilde{G}_{2,0}(k^+_1, -k^+_1 - k^+)$ of the periodic part of $G_2(k_1, -k_2 ; X)$.
For $k^+ = 0$ one obtains the contribution of the classical zero mode of Eq.(2.5). For $k^+
\neq 0$ this zero mode component of $\varphi_2(X)$ has non-zero matrix elements between one
particle states differing by a momentum transfer $\pm k^+$. As we shall see in this case the 
lowest order pertubative result indicates that covariance of the dispersion relation is explicit 
if momenta are  measured in units of the external momentum k. A specific analysis 
of the infinite volume limit is then necessary, thereby leading to a special form of the
invariant measure $[dX^-]$. The specification is now

\begin{eqnarray}
{1 \over \sqrt{k^+_1k^+_2}} G_2(k_1,-k_2 ; X) |_{DZM}  & = & 
{1 \over V \sqrt{k^+_1(k^+_1 + k^+)}} \bar{G}_{2,0}(k^+_1, -k^+_1 - k^+)
\delta(k^+_2 - k^+_1 - k^+) \nonumber \\
& : = & C_1(k^+_1, k^+) \delta(k^+_2 - k^+_1 - k^+),
\end{eqnarray}

with $V$ originating from the invariant measure.
It is the existence of this dynamical zero mode (DZM) which leads to non-trivial
dispersion relation and critical exponents.

\subsection{Equations of motion}

The equation of motion (EM) for the total field $\Phi(X)$ is

\begin{equation}
[4 \partial_+ \partial_- + m^2] [\phi_c(X) + \Psi(X)] + {\lambda \over 3!}[\phi_c(X) + 
 \Psi(X)]^3 = 0. 
\end{equation}

The driving term in this equation is $\phi_c(X)$. Since we are interested in properties of the 
system close to the phase transition where all quantities related to
long wave length fluctuations of $\phi_c$ (the order parameter $\phi_o$ in the limit $k \to
0$) are small, only terms linear in $\varphi_2$ will be kept, giving a set of coupled equation for
$\phi_c(X)$ and $\varphi_2$.

\begin{eqnarray}
[4 \partial_+ \partial_- + m^2] \phi_c(X) = -{\lambda \over 3!}[\phi^3_c(X) + 3 \phi_c(X)
\varphi_0{^2}(X)  \nonumber
\end{eqnarray}
\begin{equation}
 + \varphi_0{^2}(X)\varphi_2(X) + \varphi_2(X)\varphi_0{^2}(X) + 
\varphi_0(X)\varphi_2(X)\varphi_0(X)],
\end{equation}\

\newpage

\begin{eqnarray}
[4 \partial_+ \partial_- + m^2] \varphi_2(X) = -{\lambda \over 3!}[ 3 \phi_c(X)\varphi_0{^2}(X)
+ \varphi_0{^2}(X)\varphi_2(X)  \nonumber
\end{eqnarray}
\begin{equation}
 + \varphi_2(X)\varphi_0{^2}(X) + \varphi_0(X)\varphi_2(X)\varphi_0(X)].
\end{equation}\

Denoting the vacuum state as $|0>$ such that $a(k^+) | 0> = 0$, $k^+ > 0$, matrix
elements of EM are taken between different states :
$<0 | EM | 0> ; <q^+_1 | EM | q^+_2> ; < 0 | EM | q^+_1 q^+_2 >$ and $<q^+_1 q^+_2 | EM |
0>$ with $| q^+_1> = a^+(q^+_1) | 0>$ etc...Eqs (2.9,2.10) give then a coupled system for
$\phi_c(X) , g^{++}_2$ and $G_2$. The derivation is straightforward but tedious. We only give
the results for the Fourier components, $\tilde{\phi}_c, \tilde{g}^{++}_2$ and $ \tilde{G}_2$,
with a renormalized squared-mass $\mu^2$

\begin{equation}
\mu^2 = m^2 + {\lambda \over 8\pi} \int^{\infty}_0  {dk^+ \over k^+} \hat{f}^2(k^+), 
\end{equation}\

coming from $\varphi_0{^2}(X)$ contributions present in the right hand side of Eqs.(2.9,2.10):

\begin{eqnarray}
[\mu^2 - k^+k^-] \tilde{\phi}_c(k^+,k^-) + {\lambda \over 6 \pi} \int {dk^+_1 dk^+_2
\over k^+_1k^+_2} \theta(k^+_1) \theta(k^+_2) \hat{f}^2(k^+_1)  \hat{f}^2(k^+_2) \nonumber 
\end{eqnarray}
\begin{equation}
[{1 \over 4} \tilde{G}_2 (k^+_1,-k^+_2 ; k) + \tilde{g}^{++}_2(k^+_1, k^+_2 ; k)] = 0,
\end{equation}\

\begin{eqnarray}
[\mu^2 - (q_1 - q_2 + k)^2] \tilde{G}_2 (q^+_1,-q^+_2 ; k) + {\lambda \over 8 \pi}
\int {dk^+_1 \over k^+_1} \theta(k^+_1) \hat{f}^2(k^+_1) \nonumber
\end{eqnarray}
\begin{eqnarray} 
[\tilde{G}_2 (k^+_1,-q^+_2 ; k)  + \tilde{G}_2 (q^+_1,-k^+_1 ; k) 
+ 2 \tilde{g}^{++}_2(k^+_1, q^+_1 ; k) + 2 \tilde{g}^{++}_2(k^+_1, q^+_2 ; k)] \nonumber 
\end{eqnarray}
\begin{equation}
+ {\lambda \over 4 \pi}  \tilde{\phi}_c(k^+,k^-) = 0, \end{equation}\\

\begin{eqnarray}
 [\mu^2 - (q_1+q_2+k)^2] \tilde{g}^{++}_2(q_1^+,q_2^+ ; k) + {\lambda \over 16 \pi}
\int {dk^+_1 \over k^+_1} \theta(k^+_1) \hat{f}^2(k^+_1) \nonumber
\end{eqnarray}
\begin{eqnarray}
[\tilde{G}_2 (q^+_1,-k^+_1 ; k) + \tilde{G}_2 (q^+_2,-k^+_1 ; k) + 
2\tilde{g}^{++}_2(k^+_1, q^+_1 ; k) + 2 \tilde{g}^{++}_2(k^+_1, q^+_2 ; k)] \nonumber
\end{eqnarray}
\begin{equation}
+ {\lambda \over 8 \pi} \tilde{\phi}_c(k^+,k^-) = 0. \end{equation}

It is clear that the driving term for
$\tilde{G}_2$ and $\tilde{g}^{++}_2$ is $\tilde{\phi}_c(k^+, k^-)$. Next a closer inspection of these
 equations shows that it is the X-dependence of $g_2$ (cf.
Eq.(2.2)) which induces fluctuations of the order parameter $\phi_c$. In turn this X-dependence
has its origin in the driving term $\tilde{\phi}_c$ present in Eqs.(2.12-2.14). To facilitate the
analysis further it is useful to separate explicitely the zero-mode contribution of $\varphi_2(X)$
to Eqs.(2.12-2.14). However one has to distinguish between the two cases $k^+ = 0$ and $k^+ \ne 0$.
The first case yields the equation for critical coupling and the second the covariant dispersion relation.

\setcounter{equation}{0}

\section{Solutions of the coupled equations at $k^+ = 0$. Critical coupling}

\subsection{Equations of motion}

At $k^+ = 0$ the explicit zero mode contributions to Eqs.(2.12-2.14) are obtained with the prescription
of Eq.(2.5), with the notation $C(g,q^+)$ instead of $C_0(g,q^+)$. Using

\begin{equation}
\mu^2 - (q_1 \pm q_2)^2 = \mu^2[1 - (q^+_1 \pm q^+_2)({1 \over q^+_1} \pm {1 \over q^+_2})] = \mp
\mu^2 {(q^{+2}_1 + q^{+2}_2 \pm q^+_1 q^+_2) \over q^+_1 q^+_2}, 
\end{equation}

and with dimensionless momenta (in units of $\mu$), Eqs.(2.13-2.14) may be reduced to
 $(g = {\lambda \over 4 \pi \mu^2})$

\begin{eqnarray}
\tilde{G}_2(q_1,-q_2) & = & 2 g_{20} (q_1, -q_2) - {g \over 2} ({q_1 q_2 \over q^2_1 + q^2_2 -
q_1 q_2}) \int {dk_1 \over k_1} \hat{f}^2(k_1) [2 \tilde{g}^{++}_2 (k_1, q_1) \nonumber\\
& + & 2 \tilde{g}^{++}_2 (k_1, q_2) + \tilde{G}_2(q_1,- k_1) + \tilde{G}_2(k_1,- q_2)],
\end{eqnarray}

\begin{eqnarray}
\tilde{g}^{++}_2(q_1,q_2) & = & g_{20} (q_1, q_2) + {g \over 4}({q_1 q_2 \over q^2_1 + q^2_2 +
q_1 q_2}) \int {dk_1 \over k_1} \hat{f}^2(k_1)[2 \tilde{g}^{++}_2 (k_1, q_1) \nonumber\\
& + & 2 \tilde{g}^{++}_2 (k_1, q_2) + \tilde{G}_2(q_1,- k_1) + \tilde{G}_2(q_2,- k_1)].
\end{eqnarray}

Here, for ease of notation, all the variables $(q_i, k_i)$ stand for $(q^+_i, k^+_i)$ and the source function $g_{20}$ is defined as

\begin{equation}
g_{20} (q_1, q_2) = {g \over 4} ({q_1 q_2 \over q^2_1 + q^2_2 +q_1q_2}) 
[\hat{f}^2(q_1) C(g,q_1) + \hat{f}^2(q_2) C(g,q_2) + 2 \phi_0].
\end{equation}

The presence of the zero mode in the source term $g_{20}$ of Eqs.(3.2,3.3) is now explicit through the
coefficients $C(g,q_1), C(g,q_2)$.

\subsection{Constraints}

The projection of the EQM (2.9,2.10) onto the one particle and vacuum sectors yields the constraint $\theta_3$ with
matrix elements $<q^+_1 | \theta_3| q^+_1>$ and $<0 | \theta_3 | 0>$. Their vanishing give
for the first one an equation to determine $C(g,q_1)$ in terms of $\tilde{g}^{++}_2$ and
$\tilde{G}_2$ 

\begin{eqnarray}
C(g,q_1)[q_1 + g \hat{f}^2(q_1)] & + & 2g \int {dk_1 \over k_1} \hat{f}^2(k_1) [\tilde{g}^{++}_2 (k_1, q_1) + {1
\over 4}
(\tilde{G}_2(k_1,- q_1)  + \tilde{G}_2(q_1,- k_1))]  \nonumber\\
& + & g \phi_0 = 0 , 
\end{eqnarray}

and for the last one the equation (cf Eq.(2.12)) which fixes the critical coupling once
 $\tilde{g}^{++}_2, \tilde{G}_2$ and $C$ are known. It reads

\begin{equation}
\phi_0 + {2g \over 3} \int {dk_1dk_2 \over k_1 k_2} \hat{f}^2(k_1)\hat{f}^2(k_2)
[\tilde{g}^{++}_2( k_1,k_2) + {1 \over 4} \tilde{G}_2(k_1,-k_2)] + {g \over 6} \int{dk_1 \over
k_1} \hat{f}^4(k_1) C(g,k_1) = 0.
\end{equation}

\subsection{Iterative procedure}

The starting point is the analysis of Ref \cite{GrUlWe} in which the terms in $\tilde{g}^{++}_2$ and
$\tilde{G}_2$ in Eq.(3.5) are disregarded. The properties of $\hat{f}^2(q_1)$ permit to write the lowest order solution $C^{(0)}$ as
\begin{equation}
C^{(0)} (g,q_1) = - {g \phi_0 \over (q_1 + g)}.
\end{equation}

This fixes the initial source function $g^{(0)}_{20}(q_1, q_2)$ to be used in Eqs.(3.2-3.3). With 
$\hat{f}^2(q_1) = \theta(q_1 - {\mu \over \Lambda}) - \theta ({\Lambda \over \mu} - q_1)$, where 
$\theta(x)$ is the Heaviside step function, by direct computation after two iterations in Eqs.(3.2
-3.3) it is seen that $\tilde{g}^{++}_2$ and $\tilde{G}_2$ keep the same shape as
$g^{(0)}_{20}(q_1,q_2)$ and $2 g^{(0)}_{20}(q_1,-q_2)$ respectively. The full solution is thus
seeked under the form of $g_{20}$, Eq.(3.4), with however $C(g,q_i)$  replaced by $C^{(0)}(g_f,q_i)$  where $g_f$ is an effective coupling to be determined. There are two reasons behind this approach:\\

i) with (3.4) and (3.7) the results after one iteration for $\tilde{g}^{++}_2$ and $\tilde{G}_2$ 
are given in closed forms. One finds

$$ \lim_{q_1 \to \infty} \tilde{g}^{++(1)}_{2}(q_1,q_1) = {1 \over 6} g \phi_0 - {g^2
\phi_0 \over 6} [1 - g {\pi \sqrt{3} \over 9}] {1 \over q_1} + 0 ({1 \over q^2_1}),$$

which is just $g_{20}(q_1, q_1)$, Eq.(3.4), in the same limit with an effective coupling
$g_f$ for $C^{(0)}(g_f,q_i) :$ $g_f$ = $g(1 - g{\pi \sqrt{3} \over 9}).$

Successive iterations build up the alternating sign infinite series in powers of $g {\pi \sqrt{3} \over 9}$ so that one expects  
$g_f$ to be closed to $g[1 + g{\pi \sqrt{3} \over 9}]^{-1}$, the departure coming from the coupling
of Eqs.(3.2) and (3.3).\\

ii) the full solution for $C(g,q_1)$ from Eq.(3.5) is not expected to change drastically
from the lowest order result of Eq (3.7). Indeed using in Eq (3.5) for 
 $\tilde{g}^{++}_2$ and  $\tilde{G}_2$  the ansatz build out of $g_{20}$ with a not yet
known coefficient $C(g,q)$ one finds

\begin{equation}
C(g,q) = - {g \phi_0 \over (g+q)} - {g^2 q \over 2(g+q)} \int^{\infty}_0 dk C(g,k) [{1
\over (k^2+q^2+kq)} - {1 \over 2} {1\over (k^2+q^2-kq)} ].
\end{equation}

To a good approximation one may write

$$ \int^{\infty}_0 dk {C(g,k) \over (k^2+q^2 \pm kq)} \sim C(g,q) \int^{\infty}_0 {dk \over (k^2+q^2 
\pm kq)}
= C(g,q) { \pi \sqrt{3}\over 9 q}( 3 \mp 1).$$

Hence with this approximation the integral term in Eq.(3.8) is exactly zero. In any case
$C(g,q)$ will remain close to its lowest order form ({\it cf.} Appendix A for details).

To take into account the coupling between Eqs (3.2) and (3.3) the effective coupling
$g_f$ is written under the form

\begin{equation}
g_f = {g \over (1 +g {\pi x(g) \sqrt{3} \over 9})}. 
\end{equation}

The function $x(g)$  is expected to be close to 1 and 
is determined by  a least-square optimisation over a set of grid
points $\lbrace q_i, q_j \rbrace$ of the solution of Eqs. (3.2) and (3.3) with the ansatzs 
for $\tilde{g}^{++}_2$ and $\tilde{G}_2$ as explained above.
(cf. Appendix A). For $0 \leq g \leq 7$ it is found that a good fit to numerical values is
obtained with 

\begin{equation}
x(g) = 1 + {g \over 12}.
\end{equation}

The critical coupling $g_{c}$ is finally obtained by evaluating Eq.(3.6). With
$\hat{f}^2(k)$ given in term of step functions all the integrals are obtained in closed
forms. Divergences are all proportional to $\int^{\infty}_0 {dk \over k} 
\hat{f}^2(k)$ and are taken care of by a mass-type counter-term as discussed in Ref\cite{GrUlWe}.
The equation for $g_{c}$ is found as

\begin{equation}
1 - {g \over 6} ( 1 + {2 \over 9} g \pi \sqrt{3}) \ln [g] + {1 \over 27} g^2 \pi \sqrt{3} \ln
[1 + {1 \over 9} g \pi \sqrt{3} (1 + {g \over 12})] = 0. 
\end{equation}

To lowest order in the Haag expansion the equation was just $1-{g \over 6} \ln(g) = 0$ with 
the solution $g^{(0)}_{c} = 4.2...$. Substracting the new equation from the zeroth order one gives
a small and always negative quantity increasing with $g$, indicating that  $g_c > 4.2$. The solution of Eq.(3.11) to second order around $g^{(0)}_c$ is $g_c = 4.76...$ while the direct numerical
evaluation gives $g_c = 4.78$. 
 For comparison with other studies using the convention of Parisi et al \cite{Parisi}
 this has to be translated \cite{GrUlWe} into the reduced coupling unit $r$ which is just 1 at the perturbative
one loop level. To the above values of $g^{(0)}_{c}$ and $g_{c}$ corresponds the values $r = 1.5$ and
$r = 1.71$. This is in agreement with recent estimates from  high-temperature ($r=1.754$) \cite{BuCo,PeVi} and
 strong coupling expansions ($r=1.746$) \cite{CaPe} and  Monte Carlo simulations ($r=1.71$) \cite{KiPa,Kim}.
 However the recent RG-improved fifth order perturbative result of Ref. \cite{OrSo} is $r_5 = 1.837(30) $. A recent
high precision estimate \cite{CaHa} of $g_4$ for the 2D Ising model leads to $r = {3 g_4 \over 8 \pi} = 1.7543...$. The
discussion on the peculiarity of the perturbative result is postponed to the paragraph on critical exponents.

\setcounter{equation}{0}

\section{ Dispersion relation, critical exponents and $\beta$-function}

\subsection{dispersion relation}

Next we discuss the dispersion relation for the order parameter fluctuations, {\it i.e.}
for the field $\phi_c(X)$. Near the phase transition the equations of motion (2.12-14) are dominated 
by the dynamical zero modes. As a first step, in keeping with the treatment of Eq.(3.5), the lowest order solution is obtained by disregarding the non-zero mode contributions to Eq.(2.13). With the expressions of Eq.(2.7) for $G_2$, Eq.(2.13) reduces to

\begin{equation} 
\mu^2\bar G_{20}(q_1^+,-q_1^+-k^+) + {\lambda \over 8 \pi V}\Big[{f^2(q_1^+) \over q_1^+}+{f^2(q_1^++k^+) \over q_1^++k^+}\Big] \bar G_{20}(q_1^+,-q_1^+-k^+)+{\lambda \over 4 \pi} \phi_c(k) =0,
\end{equation}

and, with the properties of $f^2(q_1^+)$,

\begin{equation}
C_1^{(0)}(g,q_1^+,k^+) = -{g\phi_c(k)\sqrt{(q_1^++k^+)q_1^+} \over V(q_1^++k^+)q_1^+ + {g \over 2}(2q_1^++k^+)}.
\end{equation}
Going back to Eq.(2.12) for $\phi_c(k)$ the DZM contribution shows up explicitely

\begin{eqnarray}
[\mu^2-k^2]\phi_c(k) & + & {\lambda \over 24 \pi V} \int_0^{\infty}dk_1^+{f^2(k_1^+)f^2(k_1^++k^+) \over k_1^+(k_1^++k^+)}\bar G_{20}(k_1^+,-k_1^+-k^+) \nonumber\\
 & + & \mbox {terms coming from non-zero modes} = 0.
\end{eqnarray}

$\bar G_{20}$ being proportional to $\phi_c(k)$ this DZM term, irrespective of other contributions due
to corrections arising from the coupling of $g_2^{++}$ with $\bar G_{20}$ through Eq.(2.14), is therefore the first non trivial correction to the dispersion relation of the non-interacting system (on-shell condition).\\
With Eq.(4.3) as it stands the question of the covariance of the DZM contribution immediately arises. To examine this issue it is useful to recall how covariance emerges in the perturbative treatment with LC variables. Without the zero-mode the solution for $\tilde G_2$ and  $\tilde g_2^{++}$  to lowest order in $\lambda$ is obtained from  Eqs.(2.13-14) as

\begin{equation}
\tilde G_{20}(q_1,-q_2;k)=-{\lambda \over 4 \pi}{\phi_c(k) \over (\mu^2-(q_1-q_2+k)^2)},
\end{equation}

and

\begin{equation}
\tilde g_2^{++}(q_1,q_2;k)=-{\lambda \over 8 \pi}{\phi_c(k) \over (\mu^2-(q_1+q_2+k)^2)}.
\end{equation}

\par

Using these expressions in Eq.(2.12) the corresponding contribution is shown in Fig. 1\\
\vspace*{0.5cm}
\par
\vspace{-0.5cm}\hspace*{6.6cm}$k^+-k_1^+-k_2^+$\\
\par
\vspace{0.5cm}\hspace*{4.5cm}$k^+$\hspace*{1.5cm}$k_2^+$\\
\par
\vspace{0.5cm}\hspace*{6.5cm}$k_1^+$\\

\vspace{-3.0cm}\hspace*{4.2cm}{\psfig{file=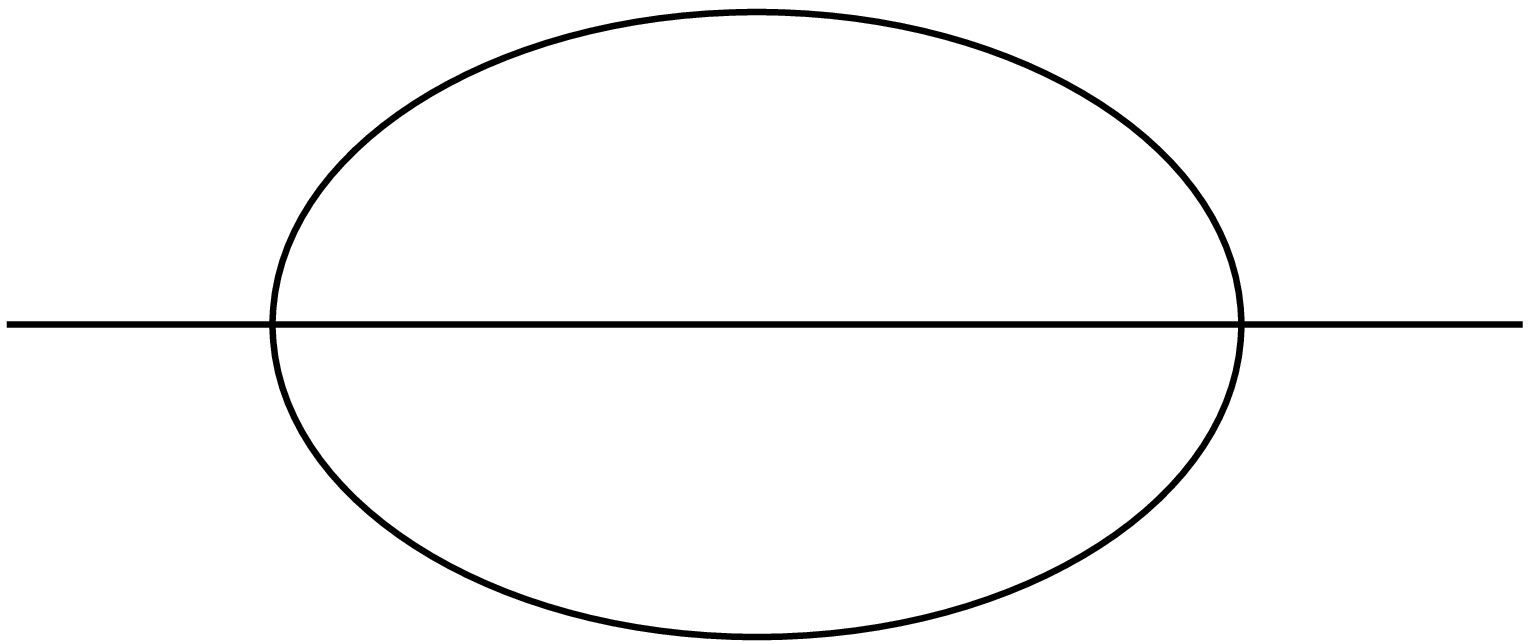,height=3cm,width=6cm}}

\vspace{0.2cm}\hspace*{2.0cm}{FIG. 1. Perturbative contribution to the self-energy at 2nd order in $\lambda$ }\\

\par

With LC variables it reads

\begin{equation}
I(k^+,k^-)=-{\lambda^2 \over 96 \pi^2} \int_0^{k^+}{dk_1^+ \over k_1^+}\int_0^{k_1^+-k^+}{dk_2^+ \over k_2^+}{f^2(k_1^+)f^2(k_2^+) \over (k^+-k_1^+-k_2^+)(k^--{\mu^2 \over k^+-k_1^+-k_2^+}-{\mu^2
\over k_1^+}-{\mu^2  \over k_2^+})}.
\end{equation}
Evaluating all momenta in units of $k^+$, that is $k_1^+=x k^+, k_2^+=y k^+$, $I(k^+,k^-)$ becomes evidently covariant, {\it i.e.} $I(k^+, k^-) \equiv I(k^2)$.

\begin{equation}
I(k^2)=-{\lambda^2 \over 96 \pi^2 \mu^2} \int_0^1dx\int_0^{1-x}dy{f^2(x)f^2(y) \over [{k^2 \over \mu^2}xy(1-x-y)-xy-(x+y)(1-x-y)]}.
\end{equation}

Thus in Eq.(4-2), with $q_1^+$ expressed in units of $k^+$, the volume $V$, which has dimension of length, must also be written in units of $1 \over k^+$, {\it i.e.} V is just ${1 \over k^+}$ times a function of  Lorentz scalars. What is its argument? The external source function $\phi_c(X)$ ({\it cf}
Eqs.(2.9-10)) actually serves to probe characteristic distances of the system. For a periodic 
background field the interaction provokes a dispersion which induces another length scale, the energy
 flow scale, which is related to the off-shellness of the process. It is measured in terms of $k^-$, 
say $k^- \over \mu^2$. Hence in effect $V  \propto  ({1 \over k^+})^{\alpha}({k^- \over \mu^2})^{(1-\alpha)}={1 \over k^+}({k^2 \over \mu^2})^{(1-\alpha)}$. Since $\alpha$ is arbitrary any scalar function 
of ${k^2 \over \mu^2}$ is legitimate. Thus 

\begin{eqnarray}
V  & = & {1 \over k^+} {\it v}({k^2 \over \mu^2}) \nonumber\\
   & \equiv & {k^- \over \mu^2} {\it w}({\mu^2 \over k^2}),
\end{eqnarray}

with ${\it v}(z) = z {\it w}(1/z)$. The infinite volume limit is then performed such that $(k^+ \to 
0, k^- \to \infty)$  with $k^2$ fixed. It is interesting to note that the two forms of $V$ in Eq.(4.8) are related 
by a conformal inversion $k^+ \to \mu^2 k^+/k^2, k^- \to \mu^2 k^-/k^2$. We shall come back
to the specific LC transformation properties susceptible to fix further the explicit form of ${\it v}(z)$.\\

With Eq.(4.8) the DZM coefficient $C_1^{(0)}$ of Eq.(4.2) takes now the following form, 

\begin{equation}
C_1^{(0)}(g,q_1^+ = x k^+,k^+) \equiv C_1^{(0)}(g,x,k) = -{g \phi_c(k) \sqrt{x(1+x)} \over {\it v}({k^2 
\over\mu^2})x(1+x) + {g \over 2}(2x+1)}.
\end{equation}

The zero modes covering the whole system, the presence of the volume in $C_1^{(0)}$ of Eq.(4.2) is not 
surprising. For a continuum formulation the finiteness of $C_1^{(0)}$ in the infinite volume limit is not 
evident at first sight. In the standard discretized approach (DLCQ) all momenta are inversely proportional 
to the size $L$ of the compact domain, indicating that $C_1^{(0)}$ should be finite in the infinite $L$ 
limit. However the restauration of covariance in this case is not clear \cite{Trento}.  It is plain to see that in the continuum approach the infinite volume limit can be taken in a 
consistent way which, in keeping with the perturbative approach, restores also the covariance of the DZM 
expressions.\\

Going back to Eq.(4.3) the DZM contribution can be written as $I_{DZM}(k^2)\phi_c(k)$ 

\begin{eqnarray}
I_{DZM}(k^2) \phi_c(k) & = & -{\lambda  \over 24 \pi} \int_0^\infty {dx \over \sqrt{x(1+x)}} C_1^0(g,x,k) \nonumber\\
 & = &  -{\lambda g \over 24 \pi} {\phi_c(k) \over \sqrt{g^2+{\it v}^{2}({k^2 \over\mu^2})}} 
\ln \Big[{g+{\it v}({k^2 \over\mu^2})+\sqrt{g^2+{\it v}^{2}({k^2 \over\mu^2})} \over g+{\it v}({k^2 \over\mu^2})-\sqrt{g^2+{\it v}^{2}({k^2 \over\mu^2})}}\Big].
\end{eqnarray}

The dispersion relation becomes simply

\begin{equation}
k^2-\mu^2-I_{DZM}(k^2) = 0.
\end{equation}

In the limit $k^2 \to 0$ it should reproduce the constraint (after renormalization)
 
\begin{equation}
\theta_3= \phi_c \mu^2(1-{g \over 6} \ln (g)) = 0.
\end{equation}

The comparaison of these last two relations shows that $\lim_{k^2 \to 0} {\it v}({k^2 \over \mu^2})$
should be zero. For ${\it v}({k^2 \over \mu^2}) \ll g$ the following expansion 
holds 

\begin{equation}
{1 \over \mu^2} I_{DZM}(k^2) \Big|_{{\it v} \ll g} = - {g \over 6} [\ln (g) -\ln ({{\it v} \over 2}) +{{\it v}
\over g}] + {\it O}({\it v}^2 \ln ({\it v})).
\end{equation}

In this case the divergent part is given by the $\ln ({\it v})$ term and is taken care of by renormalisation. If 
$\lim_{k^2 \to 0} {\it v}({k^2 \over \mu^2}) \propto (k^2)^\alpha$ with $\alpha < 1$  the leading term in the
dispersion relation is the one linear in ${\it v}$ \cite{Note1}. Apparently within our approximations for the solution of the 
EQM the dependence of ${\it v}$ on $k^2$ determines completely the dispersion relation and therefore the critical
exponent $\eta$. Given that $\eta$ is sensitive to extensive properties of the system this result is not surprising.\\

To proceed further we discuss now the physical conditions possibly fixing the function ${\it v}({k^2 \over \mu^2})$.
The volume $V$ of Eq.(4.8)  should be at least stationnary under variations of $(k^+,k^-)$ preserving the local light cone
structure. The identity map for the upper-half plane $\Sigma$ with $k^+ > 0$ is given by the projective special
linear group $PSL(2,\RR)$. Let $z = k^- + I k^+$, then an automorphism of $\Sigma$ must be of the form \cite{Naka,Hat}

\begin{equation}
T(z) = {\alpha z + \beta \over \gamma z + \delta},
\end{equation}
with $\alpha,\beta,\gamma,\delta \in \RR$ and $\alpha \delta - \gamma \beta > 0$. This can always be reduced by
projection to $\alpha \delta - \gamma \beta = 1$. The conformal transformation (4.14) is the product of a
dilatation, a translation and an inversion. The translation is purely real and concerns only $k^-$ (the energy),
the origin of which is arbitrary. Since in the infinite volume limit dilatations do not change the physical situation
special conformal transformations (SCT) are then the only relevant ones to consider ({\it cf} Appendix B). SCT's corresponds to $\alpha =
1,\beta = 0,\gamma = c$ and $\delta = 1$. To derive a differential equation for ${\it v}({k^2 \over \mu^2})$ we consider the infinitesimal form of $T(z)$ 

\begin{equation}
\delta T(z) = z - \delta_c z^2.
\end{equation}
Putting $V=V(z,\bar z) $ and  $X = {k^2 \over \mu^2}$ the evaluation
of $\Delta V = V(T(z),\overline {T(z)})-V(z,\bar z)$
in the limit ($k^+ \to 0, k^- \to \infty$, $k^2$ {\it fixed}) ({\it cf.} Appendix B) , yields the differential equation: $v'(X) = {2 \over 3} {v(X) \over X}$
 ; with $v(0)=0$ the solution
becomes $v(X) = X^{({2 \over 3})}=({k \over \mu})^{4 \over 3}$  yielding $\eta = 2-4/3=2/3$ for the critical exponent. For the universality class of the Ising model the exact result is $\eta = 1/4$. Hence the simplest DZM contribution $4/3$ already accounts for about
$76 \%$ of the exact result $7/4$ . However there are corrections
to the DZM coefficient $C_1^{(0)}$ of Eq.(4.9) coming from interaction
terms neglected in Eq.(2.13). Can they provide the missing contributions? With their inclusion one can see that Eq.(4.9) becomes
\begin{equation}
C_1(g,x,k) = -{g \sqrt{x(1+x)}[\phi_c(k) + A(g,x,k)] \over {\it v}({k^2 
\over\mu^2})x(1+x) + {g \over 2}(2x+1) + 2 g x B(g,x,k)}.
\end{equation}
The calculation of the functions $A(g,x,k)$ and $B(g,x,k)$ is detailed in Appendix C. For $g = 4.2$ the coefficients
$C_1(g,x,k)$ above and $C_1^{(0)}(g,x,k)$ , Eq.(4.9), both divided by $\phi_c(k)\sqrt{x(x+1)}$, are shown in Fig. $(2.a)$ as a function of $x$ for $k = 0.1 \mu$ and in  Fig. $(2.b)$ as a function of $k$ for $x=1$.
\vspace*{-2.0cm}\\
\hspace*{0.5cm}\psfig{file=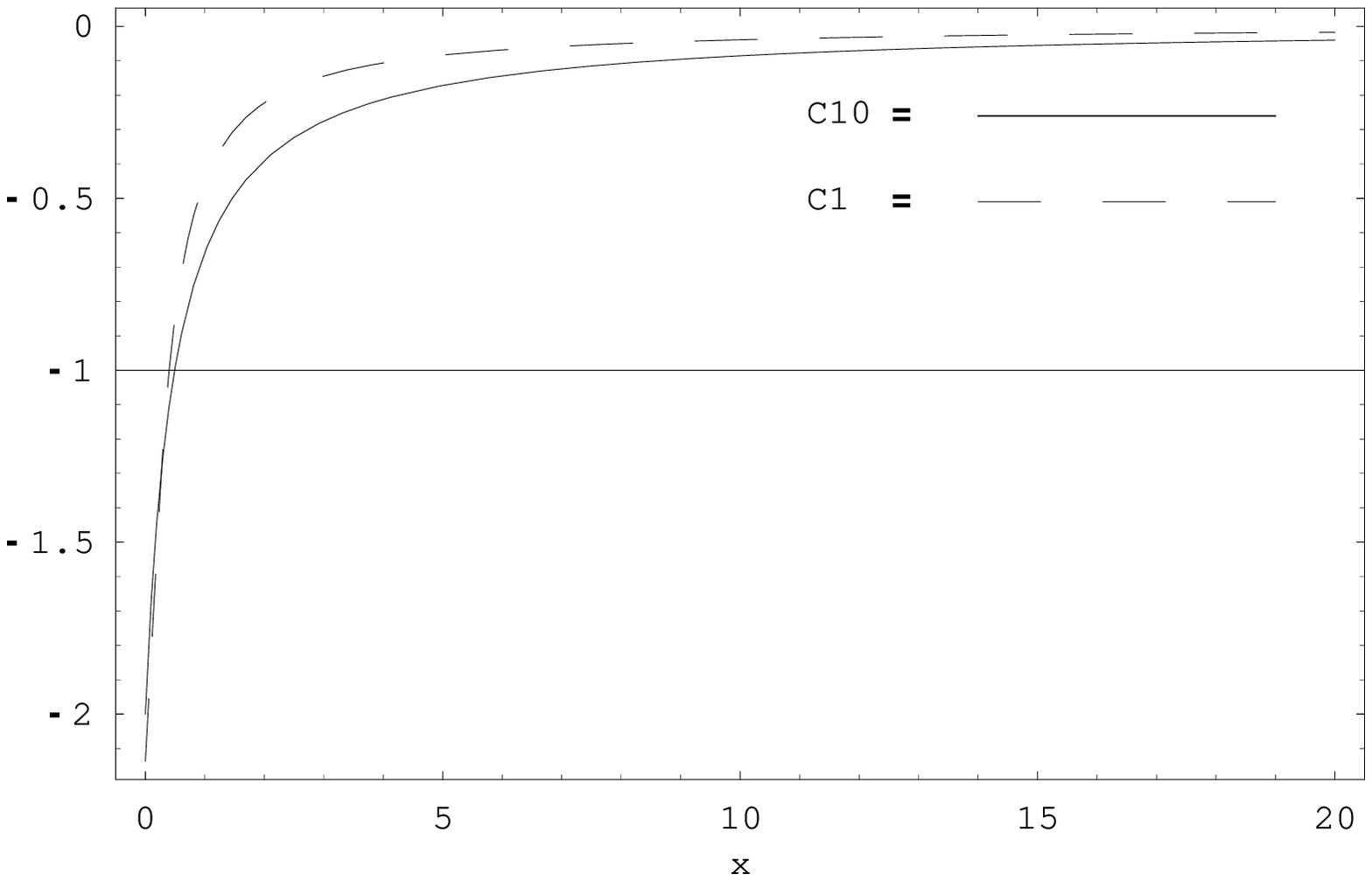,height=11cm,width=6cm}\hspace*{2.0cm}\psfig{file=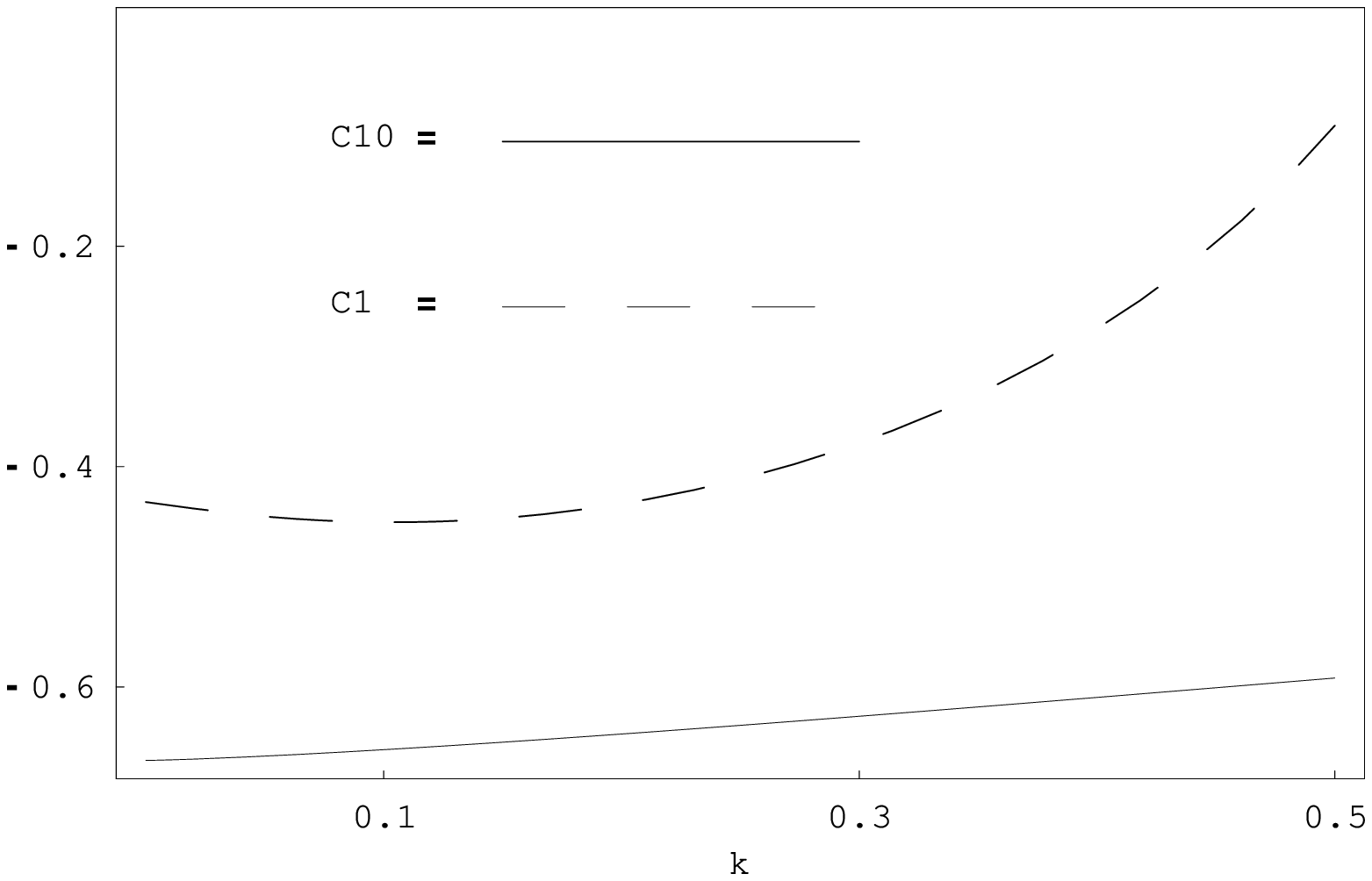,height=11cm,width=6cm}
\vspace*{-3.0cm}\\
\hspace*{3.5cm}{$(2.a)$}\hspace*{7.2cm}{$(2.b)$}\\
\vspace*{0.2cm}\\
\hspace*{1.5cm}{FIG. 2. $C_1(g,x,k)$ and $C_1^{(0)}(g,x,k)$ as a function of $x$ for $k=0.1$ $(2.a)$ and of $k$\\
\hspace*{2.7cm}for $x=1$ $(2.b)$, both for $g=4.2$. }\\

With this new value of $C_1$ the dispersion relation integral $I_{DZM}(k^2)$ can only be evaluated numerically for given values of $g$ and $k$. Fig. $(2.a)$ indicates that the result should be close to the initial one with however some difference in the $k$-dependence as seen from Fig. $(2.b)$. In the limit $k \rightarrow 0$ and for $g=4.2$ the initial dispersion integral $\tilde I_{DZM}(k^2)={24 \pi \over \lambda}I_{DZM}(k^2)$ of Eq.(4.10) reduces to ({\it cf} \, Eq.(4.13))
\begin{equation}
\lim_{k\rightarrow 0}\tilde I_{DZM}(k^2) \Big|_{C_1^0} = -2.12585+1.3333 \log (k)-0.238663 k^{({4 \over 3})}+0.046304  k^{({8 \over 3})}.
\end{equation}
The two functions $\tilde I_{DZM}(k^2)$, evaluated as above and by numerical integration with $C_1(g,x,k)$, are plotted in Fig $(3.a)$ as a function of $k$ for $g=4.2$. In the later case a least square fit ({\it cf.} Appendix C) with a functionnal of the form of Eq.(4.17) gives
\begin{equation}
\lim_{k\rightarrow 0}\tilde I_{DZM}(k^2) \Big|_{C_1} = -2.3278+1.3333 \log (k)-27.103 k^{({7 \over 4})}+3040.6  k^{({7 \over 2})},
\end{equation}
which is compared to the result of the numerical integration with ${C_1}$ in 
Fig. (3.b).
\vspace*{-2.0cm}\\
\hspace*{0.5cm}\psfig{file=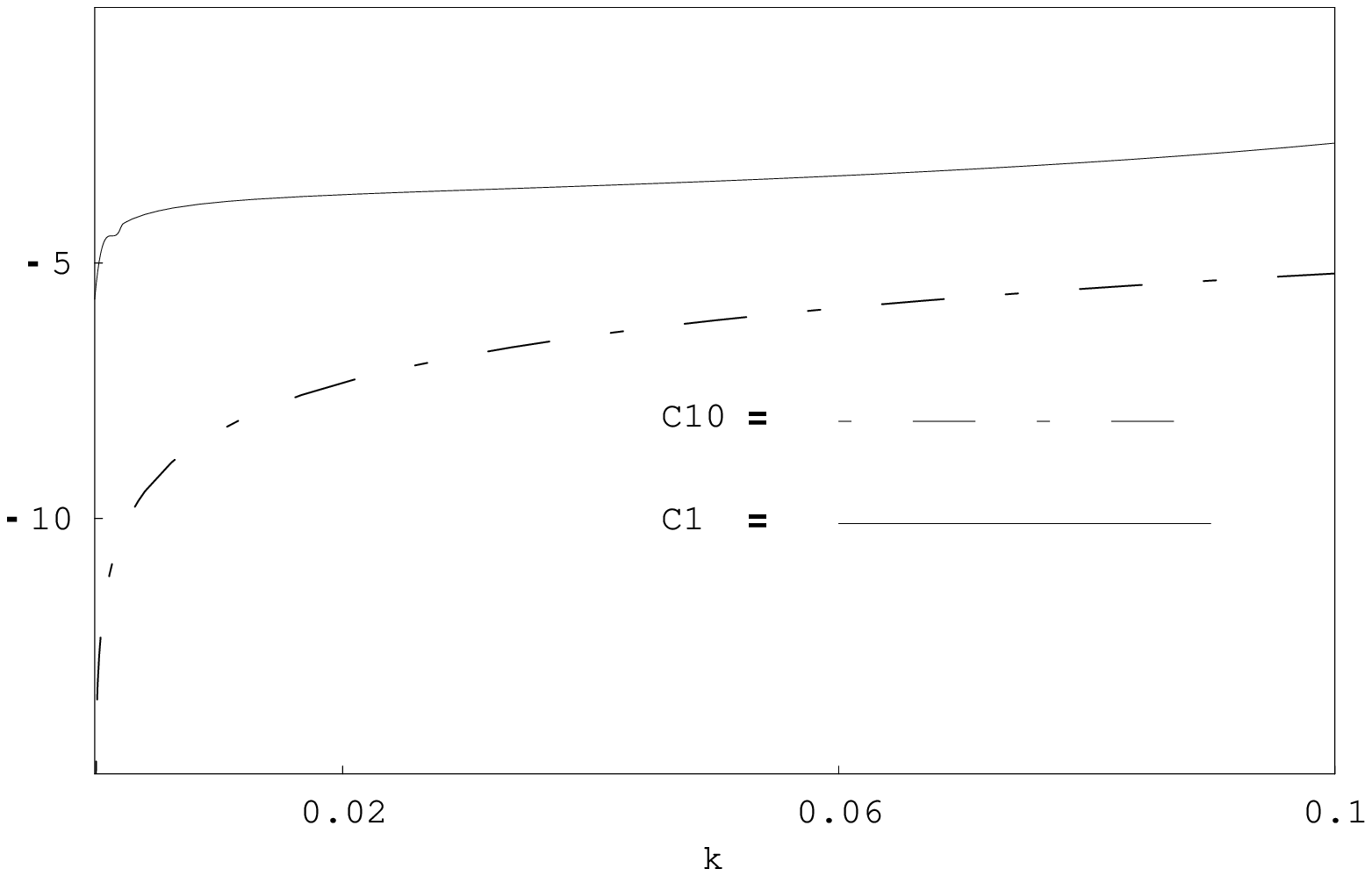,height=11cm,width=6cm}\hspace*{2.0cm}\psfig{file=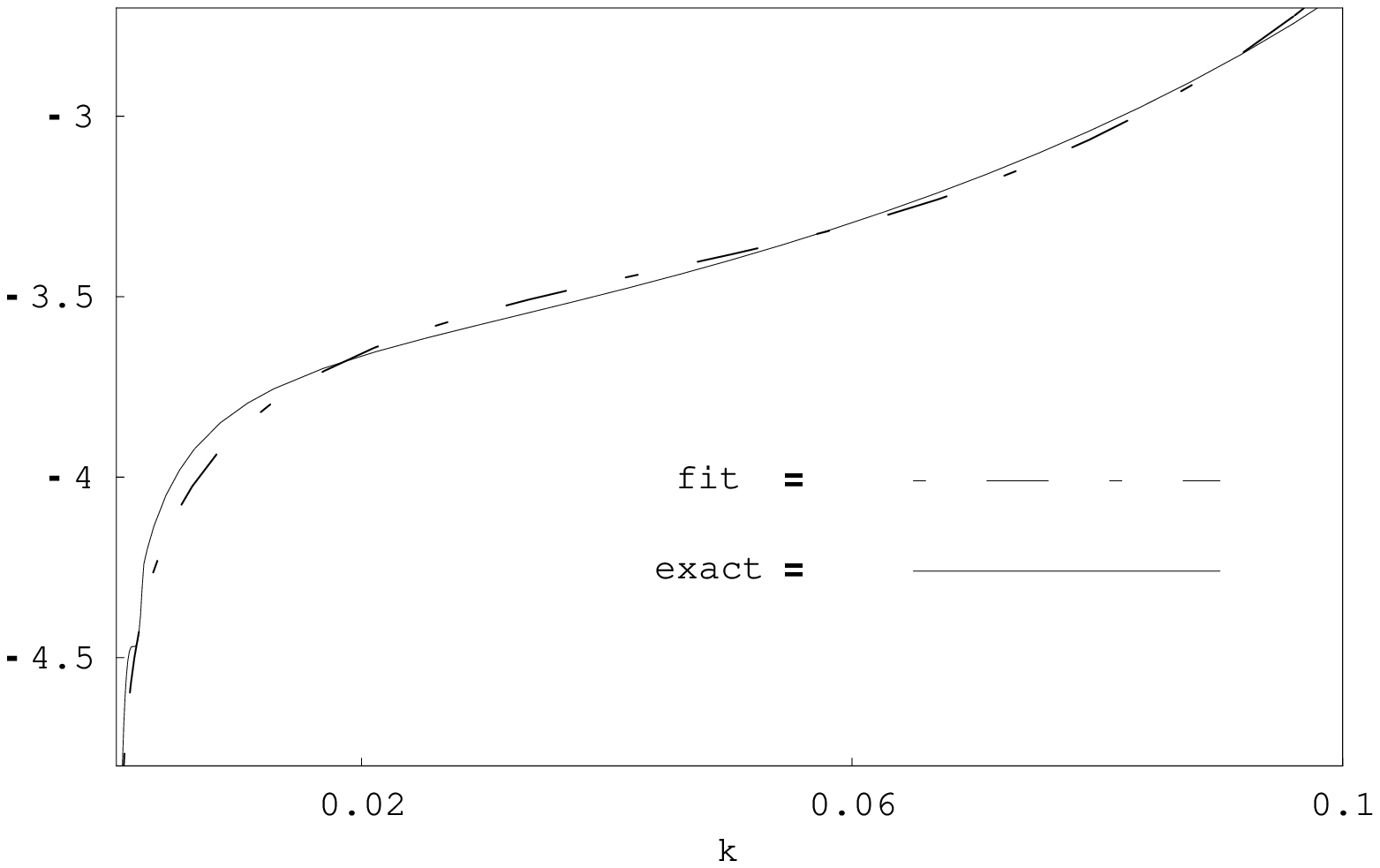,height=11cm,width=6cm}
\vspace*{-3.0cm}\\
\hspace*{3.39cm}{$(3.a)$}\hspace*{7.3cm}{$(3.b)$}\\
\vspace*{0.2cm}\\
\hspace*{1.5cm}{FIG. 3. The dispersion integrals $\tilde I_{DZM}(k^2) \Big|_{C_1^0}$ and $\tilde I_{DZM}(k^2) \Big|_{C_1}$ in $(3.a)$ and the\\
\hspace*{2.7cm} numerical and fitted $(Eq. (4.18))$ expressions in $(3.b)$ as a function of $k$  for $g=4.2$. }\\

It seems remarkable that the expected power of $k$, ${7 \over 4}$ giving $\eta =2-{7 \over 4}={1 \over 4}$, provides quite a good fit. However, despite the confidence in our fitting procedure -it has been tested
in fitting the exact integration performed with $C_{1}^{(0)}$ instead
of $C_{1}$, whereby the result (4.17) is reproduced with high accuracy ({\it cf.} Appendic C) - the least square criterion is not stringent enough to fix the
power of $k$ with total trust. E.g. a 10 \% variation of the $\chi ^{2}$ per
point results from the arbitrariness of the distribution of points between k=0
and k=0.1. This leads to a corresponding uncertainty of the power of $k$ of about
$3 \%$ which locates $\eta$ between $0.2$ and $0.3$ or in other terms $\eta=0.25 \pm0.05$. An analytical analysis of the behaviour of the full dispersion
relation at low $k^{2}$ is the only way to settle the issue without ambiguities.
Unfortunately we have not been able to perform such an analytical investigation
of the interaction corrections to the dispersion integral $I_{DZM}(k^2)$.
\subsection{ $\beta$-function}

To determine the $\beta$-function the constraint, Eq.(3.5), which depends on the cutt-off
$\Lambda$ through the test-function $\hat{f}^2(k)$, is considered as a prescription for
the calculation of the critical mass $M(g, \Lambda)$. $M^2(g, \Lambda)$ is just $\mu^2$
times the right hand side of Eq.(3.11), with $\mu^2 = \lambda/(4 \pi g)$.

One has

\begin{equation}
\beta(g) = M [{\partial M \over \partial g}]^{-1}_{(\lambda, \Lambda)}.
\end{equation}

Evaluating (4.19) gives

\begin{eqnarray}
\beta(g) & = & - 2g {N(g) \over D(g)}, 
\end{eqnarray}
with,
\begin{eqnarray}
N(g) & = & {[1 - {g \over 6}(1 + g {2\pi \sqrt{3} \over 9}) \ln(g) + g^2 
{\pi \sqrt{3} \over 27} \ln(1 + g {\pi \sqrt{3} \over 9} (1 + {g \over 12}))] 
[1 + g {\pi \sqrt{3} \over 9} (1 + {g \over 12})]}, \nonumber\\ 
D(g) & = &{(1 + {g \over 6}) [1 + g{ \pi \sqrt{3} \over 9} (1 + {g \over 12})] 
+ {g^2 \pi \sqrt{3} \over 27} (1 - { g^2 \pi \sqrt{3} \over 108})}  \nonumber\\
& + & {{g^2 \pi \sqrt{3} \over 27} [1 + {g\pi \sqrt{3} \over 9} (1 + {g \over 12})] 
\ln ({g \over (1 + g {\pi \sqrt{3} \over 9} (1 + {g\over 12}))})}. \nonumber
\end{eqnarray}

From standard renormalization group analysis one must have

$$\lim_{g \to 0} \beta(g) = - (4 - D)g + O(g^2)$$

which is indeed the case for $\beta(g)$ of Eq.(4.20). To lowest order in the Haag expansion the
corresponding $\beta$-function is just

\begin{equation}
\beta_0(g) = - {2g(1 - {g \over 6} \ln g) \over (1 + {g \over 6})}.
\end{equation}

\vspace*{-0.17cm}
\hspace*{3.5cm}\psfig{file=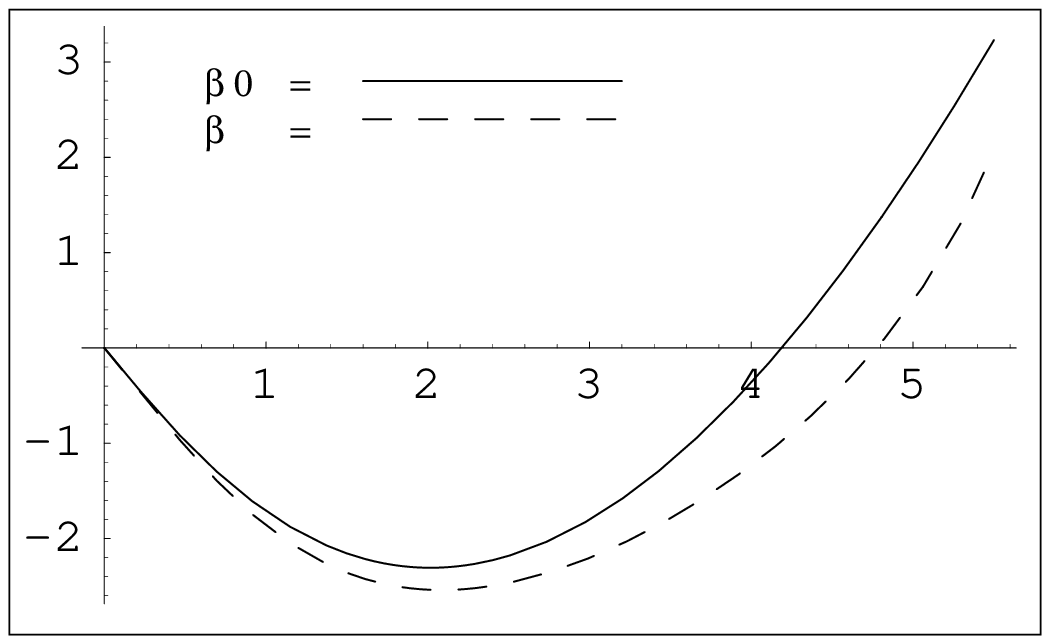,height=8cm,width=8cm}
\vspace*{-7.6cm}\\
The two functions $\beta(g)$ and $\beta_0(g)$  are plotted in Fig.4.
\vspace*{5.8cm}\\
\hspace*{5.0cm}{FIG. 4. The functions $\beta(g)$, Eqs.(4.20-4.21). }\\

In both cases the exponent $\omega=\beta^{\,\prime}(g_{cr})=2$ exactly  \cite{CCCPV}. In particular the authors of Ref\cite{CCCPV}  emphasize that at $D=2$ perturbative estimates of the critical value of $g$ and  of $\omega$  are not reliable due to strong nonanalytic corrections to the $\beta$-function. Moreover these nonanalyticities invalidate the conventionnal wisdom that $\omega$ is parametrization independant with respect to the space of coupling constants. This is most easily seen for the Ising model. With $b$ denoting the usual dimensionless Ising coupling, the critical value is $b_c={1 \over 2}\log (\sqrt{2} +1)$. Let $t=1-{b \over b_c}$, then the "mass", {\it i.e.} the inverse of the correlation length $\xi$, is a known function  of $t$, $m=t$. Using Eq.(4.20) gives $\beta(b)=-b_c t$ and $\omega=\beta '(b_c)=1$. A field theoretic definition of the coupling \cite{CCCPV} is now introduced
\begin{equation}
g^*=\lim_{t \rightarrow 0} g(t)=\lim_{t \rightarrow 0} [-{\chi_4 \over  \chi \xi^2}],
\end{equation}
 Here $\chi$ is the ''magnetic'' susceptibility, $\chi _{4}$ is
the 4-point function at zero momentum and $\xi$ the 2nd moment of the
correlation length. With this definition the new beta function becomes
\begin{equation}
\tilde \beta (g^*)=m {\partial g^* \over \partial m}=\beta(b){\partial g^* \over \partial b},
\end{equation}
which implies
\begin{equation}
\tilde \beta' (g^*)=\beta' (b)+\beta (b) {g^{*"}(b) \over g^{*'}(b)}.
\end{equation}
The common wisdom is then that $\tilde \omega=\tilde \beta' (g^{*}_{c})=\beta' (b_c)=\omega$ since
$\beta (b_c)=0$. However it is not so since $g^*(b)=t^\Delta$ with $\Delta \neq 1$.
Indeed $g^{*"}(b)/g^{*'}(b)=-(\Delta-1)/(b_c t)$ and since $\beta (b)=-b_c t$ the
correct relation is now
\begin{equation}
\tilde \omega= \omega +\Delta-1=\Delta.
\end{equation}
For the Ising model $\Delta = {7 \over 4}$, and for $\Phi^4_{1+1}$, with $g={\lambda \over 4 \pi \mu^2}$, $\omega=2$.\\
The situation was apparently known long ago \cite{And} and is examplified here in a simple and transparent way.\\

\section{Conclusion}

In the framework of LC quantization the mechanisms of symmetry breaking have
attracted a lot of interest because of their evident nonperturbative aspects.
In this respect the analysis of critical behaviour is a key issue to which,
however, little attention has been given in the past. In particular for $D=2$
it is well known that critical exponents are truely nonperturbative extensive
quantities which are not easily calculable by conventional methods. $2D$ Conformal
Field Theory (CFT) was the most spectacular achievement providing a wealth of
analytical results for a variety of models. Some specific deformations of certain
$2D$ CFT give a hint to the treatment of some particular $3D$ systems but the systematics is, so far, elusive \cite{3Dref}. 

The LC quantization scheme is an interesting nonperturbative alternative valid
in any dimension. In this paper we have shown that a full understanding of the
physics of zero modes combined with non-compact continuum field dynamics is
necessary to correctly evaluate the critical coupling and to treat covariantly the low k\( ^{+}- \) region determining the scaling behaviour. It is
found that conformal transformations preserving the local light cone structure
govern the scaling behaviour of the two-point function and provide $75 \%$ of
the total contribution to the critical exponent $\eta$ . The remaining
contributions being due to specific interaction terms were analyzed and found
to have the expected magnitude to make up for the missing $25 \%$. However, our
result for these corrections is based on a numerical analysis and should be
supported in addition by analytical investigations taking full advantage of
the conformal symmetries valid at the critical point. Thereby one could possibly
relate the CLCQ approach to the corresponding CFT. It is worth noting that the
ensemble of multicritical scalar models (the so called unitary A-A series in
CFT language) with effective Lagrangians ${\cal{L}}={1 \over 2}(\partial \varphi)^2+\varphi^{2 k}$, $k > 2$, can also be treated in CLCQ. For $k>2$
the equations of motion, the constraints and the dispersion integral  $I_{DZM}(k^2)$  will differ from the ones for  $\varphi ^4$. Comparing
the critical exponents $\eta$ with exact CFT results would provide another
test of the CLCQ DZM treatment.

\paragraph*{Acknowledgements}
E. Werner is grateful to Andr\'e Neveu for his kind hospitality
at LPM and for financial support. Discussions with A. Neveu, V. Fateev, E. Onifri,
H. de Vega and Th. Heinzl  are gratefully acknowledged. Part of this
work has been accomplished under NATO Grant $N^0 CRG920472$

\newpage
\appendix
\setcounter{equation}{0}
\Large
\centerline{\bf Appendix}
\vspace{2mm}
\small\normalsize
\section{Approximate solution of the coupled equations (3.2-3.3)}
\vspace{5mm} 
For the sake of clarity a more elaborate notation is used here. The zero order coefficient  $C_0$
of Eq.(3.7) depends also on the coupling g and on the order parameter $\phi_0$, so it is noted now 
\begin{equation}C_0(g,\phi_0,k)  =- {g \phi_0 \over (g+k)}. \end{equation}
To make clear that the source term $g_{20}$ in Eqs.(3.3-3.4) may depend on the initial and 
effective coupling $g$ and $g_f$ through the coefficient $C_0$, $g_{20}$ is noted
\begin{equation}g_{20}(g,g_f,\phi_0,q_1,q_2) = {g \over {4}} {q_1 q_2 \over (q^2_1+q^2_2+q_1 q_2)} [C_0(g_f,\phi_0,q_1)+C_0(g_f,\phi_0,q_2)+2 \phi_0].
\end{equation}
Then $g_{21}(g,g,\phi_0,q_1,q_2)$ and $G_{21}(g,g,\phi_0,q_1,q_2)$ will be the one time iterated value of
$\tilde{g}^{++}_2$ and $\tilde{G}_2$. They write explicitely
\begin{eqnarray}
g_{21}(g,g,\phi_0,q_1,q_2) & = & g_{20}(g,g,\phi_0,q_1, q_2) - g {q_1 q_2 \over (q^2_1 + g^2_2 +
q_1 q_2)}\Big[ q_1 \int^{q_1}_{0} {dk_1} {g_{20}(g,g,\phi_0,k_1, q_1) \over (k^2_1+q^2_1-k_1 q_1)}\nonumber\\
& + & q_2 \int^{q_2}_{0} {dk_1} {g_{20}(g,g,\phi_0,k_1, q_2) \over (k^2_1+q^2_2-k_1 q_2)} - {1 \over {2}} \int^{\infty}_{q_1} {dk_1 \over k_1} g_{20}(g,g,\phi_0,k_1, q_1) \nonumber\\
& - & {1 \over {2}} \int^{\infty}_{q_2} {dk_1 \over k_1} g_{20}(g,g,\phi_0,k_1, q_2)\Big]
\end{eqnarray}

\begin{eqnarray}
G_{21}(g,\phi_0,q_1,q_2) & = & 2 g_{20}(g,\phi_0,q_1, -q_2) + 2 g {q_1 q_2 \over (q^2_1 + q^2_2 -
q_1 q_2)}\Big[q_1 \int^{q_1}_{0} {dk_1} {g_{20}(g,g,\phi_0,k_1, q_1) \over (k^2_1+q^2_1-k_1 q_1)} \nonumber\\
& + & q_2 \int^{q_2}_{0} {dk_1} {g_{20}(g,g,\phi_0,k_1, q_2) \over (k^2_1+q^2_2-k_1 q_2)} - {1 \over {2}} \int^{\infty}_{q_1} {dk_1 \over k_1} g_{20}(g,g,\phi_0,k_1, q_1) \nonumber\\
& - & {1 \over {2}} \int^{q_2}_{0} {dk_1 \over k_1} g_{20}(g,g,\phi_0,k_1, q_2)\Big]. 
\end{eqnarray}
The integration limits are dictated by the fact that only states with positive momenta can be created. The integrals in Eqs.(A.3-A.4) can be obtained in closed forms. To show that $g_{20}(g,g_f,\phi_0,q_1,\pm q_2)$ with $g_f={g \over (1+g {\pi \sqrt{3} \over 9})}$ leads
to good approximate solutions to the coupled integrals equations (3.2-3.3), $g_{22}(g,g_f,\phi_0,q_1,q_2)$ and $G_{22}(g,g_f,\phi_0,q_1,q_2)$ will denote the right hand side of Eqs.(A.3) and (A.4) respectively with $g_{20}$ in the integrals replaced by $g_{20}(g,g_f,\phi_0,q_1,q_i)$.
The functions  $g_{20}(g,g_f,\phi_0,q_1,q_1)$,  $g_{21}(g,g,\phi_0,q_1,q_1)$ and $g_{22}(g,g_f ,\phi_0,q_1,q_1)$
are plotted in Fig.(A-1) as function of $q_1$ for $\phi_0=0.01$ and $g=1.5$ and $g=4.2$. 
\newpage

\vspace{-3cm}

\hspace*{1.5cm}
\psfig{file=gf15.eps,height=4cm,width=8cm}\psfig{file=gf42.eps,height=4cm,width=8cm}

\vspace{-0.05cm}\hspace*{2.5cm}{FIG. A-1. Approximate solutions of Eq.(3.3) as explained in text. }\\

 In Fig.(A-2)
the functions $2 g_{20}(g,g_f,\phi_0,q_1,-1.0)$, $G_{21}(g,g,\phi_0,q_1,1.0)$ and
 $G_{22}(g,g_f,\phi_0,q_1,1.0)$ are represented as function of $q_1$ for the same values of $\phi_0$ and $g$ (purely diagonnal cases are not relevant here).

\vspace{3cm}\hspace*{1.5cm}{\psfig{file=Gf15.eps,height=4cm,width=8cm}\psfig{file=Gf42.eps,height=4cm,width=8cm}}

\vspace{-0.05cm}\hspace*{2.5cm}{FIG. A-2. Approximate solutions of Eq.(3.2) as explained in text. }\\

 In both cases the one time iterated functions keep the same shape as the source term. Other non-diagonnal cases in $q_i$ show similar features. The agreement between 
$g_{20}(g,g_f,\phi_0,q_1,q_2)$ and $g_{22}(g,g_f,\phi_0,q_1,q_2)$ is already quite satisfactory and can be further improved for $2 g_{20}(g,g_f,\phi_0,q_1,-q_2)$ and $G_{22}(g,g_f,\phi_0,q_1,q_2)$ in the following way. The effective coupling $g_f$ is changed to

\begin{equation}
g_{f}(x)={g \over (1+g { x \pi \sqrt{3} \over 9})}
\end{equation}
 
Then a set of grid points in $q$ is chosen on which a $\chi^2$-function depending on $(g,x,\phi_0)$ is built out of the difference squared between the ansatz solutions based on $g_{20}$
,$g_{22}$ and $G_{22}$. Thus, for a given  $(g,\phi_0)$,  $x$ is the value which minimizes in a least square sense the difference between the left and right sides of the integral equations
with $\tilde{g}^{++}_{22}$ and $\tilde{G}_{22}$ replaced by the ansatz just mentionned.
The following table of values is obtained ($\phi_0=0.01$, the precise value is irrelevant)\\

\begin{tabular}{cccc|cccccccc}
 & & & g & .2 & 1 & 2 & 3 & 4 & 5 & 6 & 7 \\
\cline{4-12}
 & & & x(g) & .822 & .918 & 1.0 & 1.09 & 1.17 & 1.255 & 1.33 & 1.41\\
\end{tabular}
\vspace{5mm}

As expected $x$ remains close to one with a linear dependence $0.0835 g \sim {g \over 12}$. For $g$ small  ($ < 2$ ) the values of $x$ depend somewhat on the repartition of the grid points and the analytic argument of section (3.3) indicates that $x \cong 1$ in this cases.
For $g \le 7$ the form used hereafter is 

\begin{equation} x(g)=1+{g \over 12}. \end{equation} 

The quality of the ansatz solutions can be vievwed easily by plotting $g_{20}(g,g_f(x),\phi_0,q_1,q_2)$,\\
 $g_{22}(g,g_f(x),\phi_0,q_1,q_2)$ etc.. since all integrals in Eqs.(A.3-A.4) can be performed.

The expressions for $\tilde{g}^{++}_{22}$ and $\tilde{G}_{22}$ are now used to find the new $C(g,\phi_0,k)$, solution of Eq.(3.5). The evaluation is simple and consists in performing exactly the integrals in Eq.(3.8). For $g \le 1$ the plots do not distinguish between $C_0(g,\phi_0,k)$ of Eq.(A.1) and the new function of $k$. Their difference increases with $g$ but remains small as shown in Fig.(A.3)
for $g=5$.

\vspace{3.0cm}\hspace*{3.5cm}{\psfig{file=Ck.eps,height=5cm,width=12cm}}

\vspace{-0.0cm}\hspace*{2.0cm}{FIG. A.3. the functions $C(k)$ of Eq.(3.7) and solution of Eq.(3.5) for $g=5$ }\\

 This establishes the consistency of the procedure so that no further iteration seems needed with a more accurate expression for $C(g,\phi_0,k)$.\\

\setcounter{equation}{0}
\section{Behaviour of the volume under transformations of the projective special linear group $PSL(2,\RR)$ preserving the local light cone structure}
\vspace{5mm} 
Let $z = k^{-} + I k^{+}$. The conformal transformation $T(z) = {\alpha z + \beta \over \gamma z + \delta}$, with real coefficients $\alpha, \beta, \gamma, \delta$ such that $\alpha \delta - \beta \gamma = 1$, maps the upper-half plane onto itself and preserves the local light cone structure,
namely\\
\begin{eqnarray}
T : \left\{
\begin{array}{lll}
{\rm time like vector} & \longrightarrow & {\rm time like vector} \nonumber\\
{\rm null vector}      & \longrightarrow & {\rm null vector} \nonumber\\
{\rm spacelike vector} & \longrightarrow & {\rm spacelike vector} \nonumber 
\end{array}
\right.
\end{eqnarray} 
In full generality one may choose $\alpha = e^{\rho \over 2}(1+a c)$, $\beta = a e^{-{\rho \over 2}}$, $\gamma = c e^{\rho \over 2}$ and $\delta = e^{-{\rho \over 2}}$. T(z) results then from the successive actions of a dilatation, $z \to z e^{\rho}$, followed by a special conformal transformation (SCT), $z \to {z \over {1+c z}}$, and a translation, $z \to z+a$. For infinitesimal values of the parameters $\delta_a,\delta_c$ and $\delta_\rho$ $T(z) = z+\delta_a-\delta_c z^2+\delta_\rho z$. Requiring stationnarity  of the volume $V={1 \over k^+} v({k^2 \over \mu^2}) \equiv V(z,\bar z)$ under the infinitesimal transformation $z \to T(z)$ leads to the relation
\begin{equation}
[-{\delta_\rho \over k^-}+ 2 \delta_c] {v(X) \over X} = [{\delta_a \over (k^{-})^2}-2 {\delta_\rho \over k^-}+ \delta_c ( 3-{k^4 \over (k^{-})^4})] {\partial v(X) \over \partial X},
\end{equation}
where $X = {k^2 \over \mu^2}$. In the limit $(k^+ \to 0, k^- \to \infty)$  with $k^2$ fixed only the $\delta_c$ (SCT) contribution survives. With the initial condition
$v(0)=0$ the solution is $v(X)=X^{({2 \over 3})}$.

\setcounter{equation}{0}
\section{Corrections to the dispersion relation and power law behaviour at small $k^2$.}
\vspace{5mm} 
The starting point to evaluate corrections to  $C_1^{(0)}(g,q^+,k^+)$ of Eq.(4.2) is Eq.(2.13). Using Eq.(2.7) to extract the DZM contribution, the complete equation for $C_1(g,q^+,k^+)$ now reads
\begin{eqnarray}
& &{C_1(g,x,k) \over \sqrt{x(x+1)}}\Big[v({k^2 \over \mu^2}) x(1+x)+{g \over 2}(2 x+1)\Big] + g \phi_c(k) \\
&+&{g \over 2} \int {dz \over z} \hat{f}^2(z)\Big[\tilde G_2^{NZ}(z,-(1+x);k)+\tilde G_2^{NZ}(x,-z;k)+
2 \tilde g^{++}_2(z,x;k)+2 \tilde g^{++}_2(z,1+x;k)\Big]=0, \nonumber
\end{eqnarray}
where $q^+=x k^+$ and  $\tilde G_2^{NZ}$ is the non-zero mode contribution to $\tilde G_2$. The corrections come from the integral terms which are evaluated with the lowest order solution   $\tilde g^{++}_{20}$ and  $\tilde G_{20}^{NZ}$, {\it i.e.} obtained by neglecting integral terms in their own equation. These solutions read
\begin{eqnarray}
\tilde g^{++}_{20}(x,y;k)&=&h(x,y;k)\Big[\sqrt{{x \over 1+x}} C_1(g,x,k)+\sqrt{{y \over 1+y}} C^{(0)}_1(g,y,k)+2 \phi_c(k)\Big], \nonumber \\
\tilde G_{20}^{NZ}(x,y;k)&=&2 \phi_c(k) h(x,y;k),\,\,\,\,\,h(x,y;k)={g \over 4}\,{xy \over (1+k^2 x)y(y+x(1+x))+x(1+x)}.
\end{eqnarray}
In $\tilde g^{++}_{20}(x,y;k)$ terms which can be regrouped later on with $C_1(g,x,k)$ in Eq.(C.1) have been singled out whereas $C_1(g,y,k)$ was replaced by $C^{(0)}_1(g,y,k)$ whenever this coefficient enters under the integral to be performed. Eq.(C.1) can then be written as
\begin{equation}
{C_1(g,x,k) \over \sqrt{x(x+1)}}\Big[v({k^2 \over \mu^2}) x(1+x)+{g \over 2}(2 x+1)+2 g x B(g,x,k)\Big] + g (\phi_c(k)+ A(g,x,k))=0, \\
\end{equation}
where $A(g,x,k)=\Sigma_{i=1}^4 A_i(g,x,k)+ 2 \phi_c(k) [B(g,x,k)+B(g,1+x,k)]$ with
\begin{eqnarray}
A_1(g,x,k)&=&{1 \over 2}\int_{(1+x)}^{\infty}{dz \over z} \tilde G_{20}^{NZ}(z,-(1+x);k) \nonumber \\
A_2(g,x,k)&=&{1 \over 2}\int_{0}^{x}{dz \over z} \tilde G_{20}^{NZ}(x,-z;k) \nonumber \\
A_3(g,x,k)&=&\int_{0}^{\infty}{dz \over z} \sqrt{{z \over 1+z}} C_1^{(0)}(g,z,k) h(z,x,k) \nonumber \\
A_4(g,x,k)&=&A_3(g,1+x,k),\,\,\,\,\,\,B(g,x,k)=\int_{0}^{\infty}{dz \over z}  h(z,x,k).
\end{eqnarray}
All the integrals in $A_i(g,x,k)$ and in $B(g,x,k)$ can be obtained in closed forms but the final integral over $x$ giving the dispersion contribution $I_{DZM}(k^2)$ can only be done numerically for given values of $g$ and $k$. As seen in Fig.(2.a) $C_1(g,x,k)$ and $C_1^{(0)}(g,x,k)$ approach each other for large $x$; we therefore separate the integral over $x$ as follows
\begin{eqnarray}
\int_{0}^{\infty}{dx \over \sqrt{x(1+x)}}C_1(g,x,k)&=&\int_{0}^{50}{dx \over \sqrt{x(1+x)}}C_1(g,x,k)+
\nonumber \\
& &{C_1(g,50,k) \over C^{(0)}_1(g,50,k)} \int_{50}^{\infty}{dx \over \sqrt{x(1+x)}}C^{(0)}_1(g,x,k).
\end{eqnarray}
Thereby we avoid delicate numerical work in the limit of large $x$, since the last contribution can be evaluated analytically. For $g=4.2$ the corresponding dispersion integral $I_{DZM}(k^2)$ is shown  by the continuous curve in Fig.(3). A fitting procedure is now envisaged to extract its power law behaviour as a function of $k$ in the small $k$ region. The method is tested first on the known result with $C^{(0)}_1(g,x,k)$ which reads
\begin{equation}
\lim_{k\rightarrow 0}\tilde I_{DZM}(k^2) \Big|_{C_1^0} = -2.12585+{4 \over 3} \log (k)-0.238663 k^{({4 \over 3})}+0.046304  k^{({8 \over 3})}-0.037973 k^{({8 \over 3})} \log(k).
\end{equation}
A grid of 40 points between $k =0$ and $0.1$ is chosen according to the relation
\begin{equation}
k(i)={1 \over 2} \tan [{1 \over 2} \arctan [0.210](1+z(i))],
\end{equation}
where $z(i)$ are the 40 Gauss points with $-1 < z(i) < 1$. This choice produces points with greater density in the region where the function to be fitted is varying most rapidly. With a trial function of the form of Eq.(C.6) with unknown coefficients and powers of $k$ the constant and the coefficient of the log term are first adjusted by least-square fitting  the lower part of the spectrum.  
Thereby the domain to explore in the multiparameter space is considerably reduced. A full least-square search is then performed over the whole 40 points giving
\begin{equation}
fit(k)=-2.12585+1.33333 \log (k)-0.23848 k^{({1.3333})}+0.05382 k^{({2.6666})}
-0.03335 k^{({2.6666})} \log(k).
\end{equation}
The same tactic is used for the numerical spectrum of Fig.(3). The spectral values are given in table (C.1)\\
\par
\hspace*{-0.5cm}\begin{tabular}{cllclclc}
\begin{tabular}{cllc}
        \ $k$ & \ \ $\tilde I(k)$ \\ \\ 
	$9.12 \times {{10}^{ -5 }}$& -5.484  \\
	$4.799 \times {{10}^{ -4 }}$& -4.816  \\
	$1.177 \times {{10}^{ -3 }}$& -4.472  \\
	$2.178 \times {{10}^{ -3 }}$& -4.252  \\
	$3.477 \times {{10}^{ -3 }}$& -4.098  \\
	$5.066 \times {{10}^{ -3 }}$& -3.986  \\
	$6.937 \times {{10}^{ -3 }}$& -3.9  \\
	$9.077 \times {{10}^{ -3 }}$& -3.833  \\
	$1.147 \times {{10}^{ -2 }}$& -3.78  \\
	$1.411 \times {{10}^{ -2 }}$& -3.735  
\end{tabular}
\begin{tabular}{cllc}
        \ $k$& \ \  $\tilde I(k)$ \\ \\
	$5.396 \times {{10}^{ -2 }}$& -3.36  \\
	$5.801 \times {{10}^{ -2 }}$& -3.317  \\
	$6.204 \times {{10}^{ -2 }}$& -3.272  \\
	$6.6 \times {{10}^{ -2 }}$& -3.224  \\
	$6.99 \times {{10}^{ -2 }}$& -3.175  \\
	$7.369 \times {{10}^{ -2 }}$& -3.124  \\
	$7.735 \times {{10}^{ -2 }}$& -3.071  \\
	$8.088 \times {{10}^{ -2 }}$& -3.017  \\
	$8.423 \times {{10}^{ -2 }}$& -2.963  \\
	$8.739 \times {{10}^{ -2 }}$& -2.909 
\end{tabular} 
\begin{tabular}{cllc}
        \ $k$& \ \  $\tilde I(k)$& \\ \\
	$5.396 \times {{10}^{ -2 }}$& -3.36  \\
	$5.801 \times {{10}^{ -2 }}$& -3.317  \\
	$6.204 \times {{10}^{ -2 }}$& -3.272  \\
	$6.6 \times {{10}^{ -2 }}$& -3.224  \\
	$6.99 \times {{10}^{ -2 }}$& -3.175  \\
	$7.369 \times {{10}^{ -2 }}$& -3.124  \\
	$7.735 \times {{10}^{ -2 }}$& -3.071  \\
	$8.088 \times {{10}^{ -2 }}$& -3.017  \\
	$8.423 \times {{10}^{ -2 }}$& -2.963  \\
	$8.739 \times {{10}^{ -2 }}$& -2.909  
\end{tabular}
\begin{tabular}{cllc}
        \ $k$& \ \ $\tilde I(k)$ \\ \\
	$9.035 \times {{10}^{ -2 }}$& -2.855  \\
	$9.308 \times {{10}^{ -2 }}$& -2.802  \\
	$9.556 \times {{10}^{ -2 }}$& -2.752  \\
	$9.778 \times {{10}^{ -2 }}$& -2.705  \\
	$9.972 \times {{10}^{ -2 }}$& -2.661  \\
	$1.014 \times {{10}^{ -1 }}$& -2.623  \\
	$1.027 \times {{10}^{ -1 }}$& -2.59  \\
	$1.038 \times {{10}^{ -1 }}$& -2.563  \\
	$1.045 \times {{10}^{ -1 }}$& -2.545  \\
	$1.049 \times {{10}^{ -1 }}$& -2.534
\end{tabular}
\end{tabular}
\vspace*{-4.5cm}
\hrule
\vspace*{5.cm}

\hspace*{4.0cm}{Table C.1.  Values of $\tilde I_{DZM}(k^2) \Big|_{C_1} $ as function of $k$ \\

A least-square fit on the eight first points of the spectrum fixes the constant and the coefficient of the $log$ term to $-2.185$ and $1/3$ respectively. The remaining part  $\tilde I_{DZM}(k^2) \Big|_{C_1}+2.185-{1 \over 3} log(k)$ is shown in Fig.(C.1) and is therefore a power of k close to $2$. 

\vspace*{-2.5cm}

\hspace*{3.5cm}\psfig{file=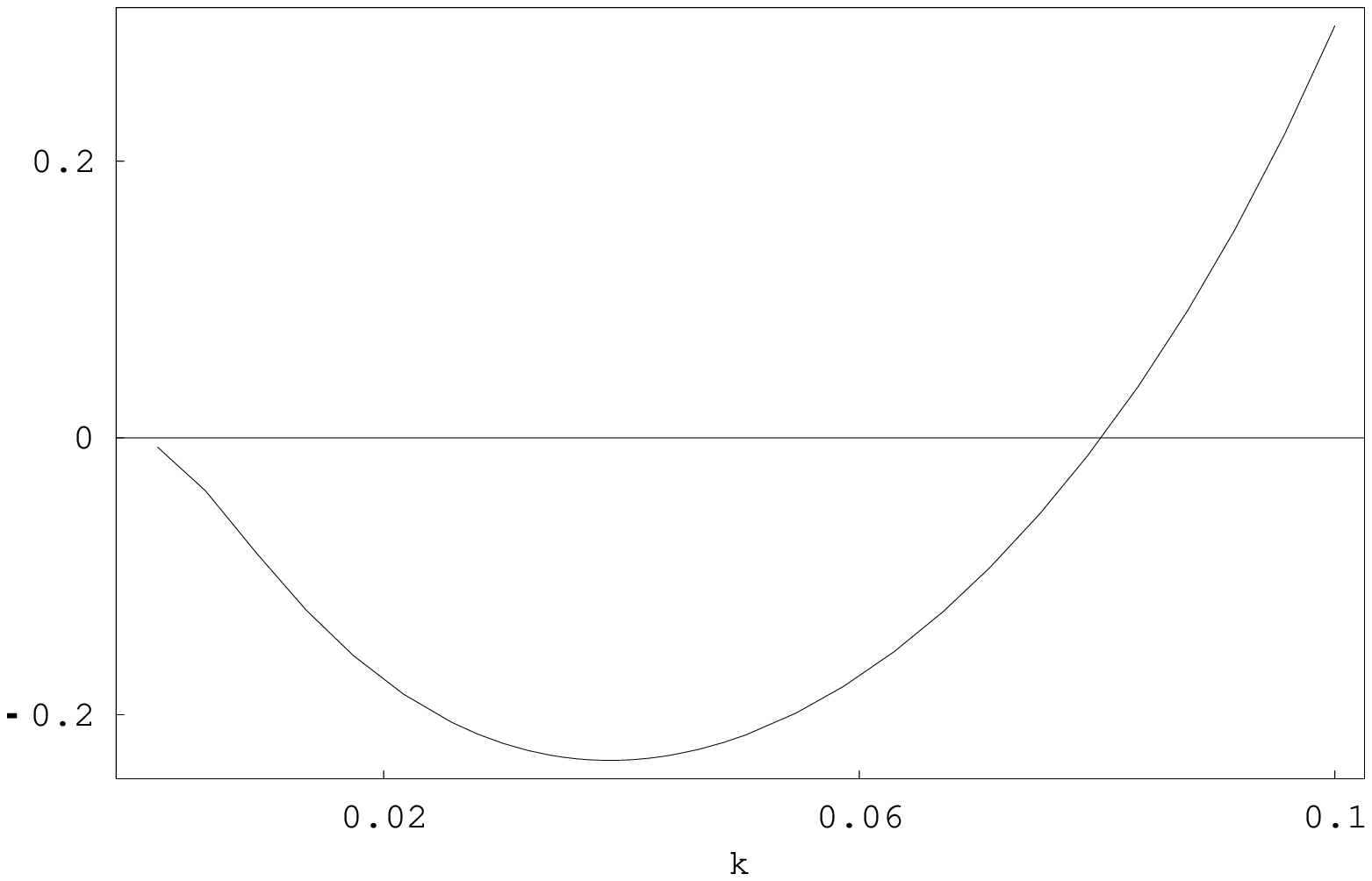,height=11cm,width=7cm}
\vspace*{-2.5cm}

\hspace*{3.cm}{FIG. C.1. $\tilde I_{DZM}(k^2) \Big|_{C_1}+2.185-{1 \over 3} log(k)$ as a function of $k$. }\\

The final least-square search with a trial fitting function $a+{1 \over 3}log (k)+ c k^e+d k^{2e}$ is made over the whole spectrum with input values $a=-2$, $c=27, e=1$ and $d=3000$ with the result  
\begin{equation}
fit(k)=-2.3278+{1 \over 3} \log (k)-26.9999 k^{({1.74959})}+3040 k^{({3.4992})}.
\end{equation}
Fixing finally $e$ to ${7 \over 4}$ gives the fit of Eq.(4.18).


\begin{thebibliography}{10}
%
\bibitem{PaBr} 
H.C.~Pauli, S.J.~Brodsky, Phys. Rev. {\bf D32}, 1993, (1985)\\ 
S.J.~Brodsky, H.C.~Pauli and S.S.~Pinsky, Phys. Report {\bf 301},229, (1998).\\
%
\bibitem{Hou}
"New Non-Perturbative Methods and Quantization on the light cone" Les Houches
Series, Vol. {\bf8} (1998), Editors : P. Grang\'e, H.C. Pauli, A. Neveu, S. Pinsky, E. Werner,
EDP-Sciences, Springer.\\
%
\bibitem{SaGrWe}
S.~Salmons, P.~Grang\'e, E.~Werner, Phys.Rev. {\bf D60}, 067701, (1999).\\
%
\bibitem{GrUlWe}
P.~Grang\'e, P.~Ullrich, E.~Werner, Phys.Rev. {\bf D57}, 4981, (1998).\\  
%
\bibitem{HKW}
T.~Heinzl, S.~Krusche, E.~Werner, Nucl. Phys. {\bf A532}, 429, (1991).\\  
%
\bibitem{Parisi}
G.~Parisi, J.Stat.Phys. {\bf 23}, 49 (1980), Nuovo Cimento {\bf
A21}, 179 (1974).\\
%
\bibitem{BuCo}
P. Butera, M. Comi, Phys. Rev {\bf B54}, 15828, (1996).\\
%
\bibitem{PeVi}
A. Pelissetto, E. Vicari, Nucl. Phys. {\bf B519}, 626, (1998).\\
%
\bibitem{CaPe}
M. Campostrini, A. Pelissetto, P. Rossi, E. Vicari,  Nucl. Phys. {\bf B459}, 207, (1996).\\
%
\bibitem{KiPa}
J.K. Kim, A. Patrascioiu, Phys. Rev {\bf D47}, 2588, (1993).\\
%
\bibitem{CCCPV}
P.~Calabrese, M.~Caselle, A.~Celli, A.~Pelissetto, E.~Vicari, J. Phys. {\bf A33}, 8155, (2000).\\
%
\bibitem{Kim}
J.K. Kim, Phys. Lett. {\bf D345}, 469, (1995).\\
%
\bibitem{OrSo}
E.V. Orlov, A.I. Sokolov, preprint hep-th/0003140 ; Fiz. Tver. Tela  {\bf42,11}, 2087, (2000).\\
%
\bibitem{CaHa}
M. Caselle, M. Hasebusch, A. Pelissetto, E. Vicari, J. Phys. {\bf A33}, 8171, (2000).\\
%
\bibitem{Trento}
 P. Grang\'e, S. Salmons, E. Werner, hep-th/0111186, Nucl. Phys. proc. sup. {\bf B108}, (2002)\\
%
\bibitem{Note1}
At D=2 all perturbative contributions to $\Gamma (k^2)$ are  analytic in $k^2$.\\
%
\bibitem{Naka}
M Nakahara, "Geometry, Topology and Physics", Graduate Student Series in Physics (1990),\\ 
Institute of Physics Publishing, Bristol and Philadelphia , D. E. Brewer Editor.\\
%
\bibitem{Hat}
B. Hatfield, "Quantum Theory of Point Particles and Strings", Addison-Wesley Publishing Company (1992)\\
%
\bibitem{And}
N. Andrei, Phys. Rev. Lett. {\bf 45}, 379, (1980).\\
%
\bibitem{3Dref}
V. A. Fatevv, preprint LPM/01-10, hep-th/0103014.\\
%
\end{thebibliography}
\end{document}